\documentstyle[12pt]{article}
\date{}
\def\be{\begin{equation}}
\def\ee{\end{equation}}
\def\ben{\begin{eqnarray}}
\def\een{\end{eqnarray}}
\def\bena{\begin{eqnarray*}}
\def\eena{\end{eqnarray*}}
\def\bdes{\begin{description}}
\def\edes{\end{description}}
\def\benum{\begin{enumerate}}
\def\eenum{\end{enumerate}}
\def\bebib{\begin{thebibliography}}
\def\eebib{\end{thebibliography}}
\def\hs{\hspace*{.3in}}
\def\hsm{\hspace{.1in}}

\def\pa{\partial}

\title{%
\vskip-7em\hfill {\small \begin{tabular}[t]{l}
                  \rule{0ex}{1ex}  \\[.0ex]
                  \rule{0ex}{1ex}
\end{tabular} \break\vskip3em  }
The Solution to Wheeler-DeWitt is Eight}

\author{\vbox{\vspace{0mm}}\\[-4mm]%
E. Adi$^1$ and S. Solomon$^{2,1}$ \\[6mm]
\small $^1\,$ Permanent address Racah Institute of Physics \\[-2mm]
\small Hebrew University, 91904 Jerusalem, ISRAEL  \\[-2mm]
\small eti@vms.huji.ac.il                          \\[-2mm]
\small sorin@vms.huji.ac.il                        \\[-2mm]
\small $^2\,$ Scuola Internazionale Superiore di Studi Avanzati \\[-2mm]
\small SISSA, Via Beirut n. 2-4,  34013 Trieste ITALY       \\[-2mm]
\small sorin@tsmi19.sissa.it                        \\[1mm]
\normalsize  {\em Submitted to Phys. Lett. B.}\\[1mm]
}
\begin{document}
\maketitle %%\vfill
%--------------------------------------------------%
%
\thispagestyle{empty}
\begin{abstract}
%\noindent \normalsize \vbox{\vspace{0mm}}\\[-4mm]

We describe a new geometrical solution of the Wheeler-
DeWitt equation in two dimensional quantum gravity.
The solution is the amplitude of a surface
whose boundary consists of two tangent loops.

We further discuss a new method for estimating singular
geometries amplitudes, which uses explicit recursive counting
of discrete surfaces.
\end{abstract}
\setcounter{page}{0}
\newpage
%------------------------------------------------%
\section{Introduction}
%What is WDW?
%How does WDW look in 2D?
%Who respects WDW in 2D?

In the quantization of gravity, wavefunctions are
functions of the spatial metric and the other fields
in the theory.
The requirement of gauge invariance
states that wavefunctions are required to obey the operator
version of the space and time diffeomorphism constraints.
The Weeler-DeWitt (WDW) equation expresses the invariance
under the generator of time-diffeomorphisms \cite{DeWitt} and
plays a fundamental role in the theory \cite{fund}.

For this reason it is very important to check
for any candidate of quantum gravity, whether it
engenders the WDW equation.

In two dimensions (2D) gravity the spatial geometry is completely
specified by the length of the 1-D boundary,
up to spatial diffeomorphisms.
Therefore the
minisuperspace approximation
(considering spatial geometries that are specified by a single
scalar factor) is exact.

In the absence of matter fields , the WDW equation in 2D becomes
\cite{moore1}:
\be\label{wdw}
(-(l\frac{\pa}{\pa l})^{2} + 4 \mu l^{2} + \nu^{2})\Psi(l)=0
\ee
Where $\Psi$ are wavefunctions describing the
the geometry of space-time.
[BThe term $\nu^{2}$ is obtained by assuming that
the states are associated to local
operators in KPZ conformal theory \cite{kpz}
and assumes positive half-integer values..

$l$ is the length of the 1-D boundary , $\mu$ is the
cosmological constant (which in the minisuperspace model
is restricted be positive) and $\nu$ is a constant that can have
positive half integer values in the case of pure gravity.

The solutions of this equation  decaying at large lengths are
the modified Bessel functions:
\be\label{solution}
\Psi_{O}(l) \sim K_{\nu}(2 \sqrt{\mu} l)
\ee
It is quite nontrivial, but true that the same result
is obtained by estimating with matrix model
techniques appropriate path integral expressions.

In particular, the path integral over all the
surfaces with one puncture,
one marked point on the boundary, and disk topology (fig.1)
was shown in \cite{moore1} to correspond to
the solution with $\nu = 1/2$ of ~(\ref{wdw}).

The physical interpretation of this integral is the
amplitude of an universe of length $l$ propagated
out of a cosmological constant insertion
(geometrically represented as a puncture).

A second solution, corresponding to
$\nu=\frac{3}{2}$  was shown to be obtained by applying the
boundary operator on surfaces with disk topology
\cite{moore1}.
The boundary operator measures the total
length of the boundaries. When applied to a disk, it
assumes the expectation value
$l w(l)$ where $w(l)$ is the amplitude for the disk.

The path integral expressions solving the equation ~(\ref{wdw})
 for higher $\nu$ values do not have such a clear interpretation.

None of the two solutions mentioned above has the original
Hawking-Hartle interpretation of being the amplitude
of creating the universe from nothing \cite{hh}:
the first one starts with an insertion , and the second one
involves
the measurement of a particular quantity (the boundary length).

The further search for purely geometric solutions to the WDW
equation
starting from  "nothing"
was  impaired by the difficulties to estimate singular
geometries amplitudes
by KPZ or matrix model techniques.

Using a new discrete counting technique ,
we compute for the first time,
the amplitude to propagate out of "nothing"
an universe with the topology of two tangent circles,
named in \cite{moore2} "figure eight diagram".

It turns out that this amplitude fulfills the WDW
equation ~(\ref{wdw})
corresponding to $\nu=\frac{3}{2}$.
We further discuss the geometrical interpretation
of this solution.

We  demonstrate the generality of our
new combinatoric method by calculating
in the last part of the letter another singular
geometry example \cite{t.b.p}
unavailable by previous techniques.
%---------------------------------------------------------%
\section{Disk amplitudes and their relation to WDW solutions}

%2) Recall what is the Disk amplitude and how does it
%relate to WDW?
%Nati's results,(KPZ,Matrix model)

The singular amplitudes computed in this paper are all related to the
regular disk amplitude. Therefore, in order to describe their
properties, we first recall some known facts on the
diagrams with disk topology in the matrix model framework.

Recall that the partition function of a disk with fixed area $A$
and perimeter $l$ has the form \cite{moore1}:
\be\label{part}
Z(A,l) = l^{\frac{1}{2}} A^{-\frac{5}{2}}
e^{-\frac {l^{2}}{A}}
\ee
The loop amplitude is defined to be:
\be\label{int}
w(l)=\int Z(A,l) e^{-\mu A}
\ee
and its value is found in \cite{moore1} to be:
\be\label{resint}
w(l)=l^{-\frac{5}{2}}e^{-2\sqrt{\mu} l}(1+2\sqrt{\mu} l)
\ee

The modified Bessel function $K_{\nu}$ admits an integral
representation with an integrand similar in form to
{}~(\ref{part}):
\be\label{bes}
K_{\nu}(z)=\frac{1}{2}(\frac{z}{2})^{\nu}\int_{0}^{\infty}dt
t^{-\nu-1}e^{-t-\frac{z^{2}}{4t}}
\hs , \hs |argz|<\frac{\pi}{2} \hs , \hs Re z^{2}>0 .
\ee
where $z = 2 \sqrt{\mu} l $ and \hsm $\nu=\mu A $.

We thus learn that ~(\ref{int}),~(\ref{resint})
can be written in terms of the modified Bessel functions as:
\be
w(l)=l^{-1}(\sqrt{\mu})^{\frac{3}{2}}K_{\frac{3}{2}}
(2 \sqrt{\mu} l)
\ee

This result for the regular disk amplitude
was found in \cite{moore1} by means of
semiclassical approximation combined with scaling
arguments, as well as by matrix models results.
However, it can also be obtained through direct counting of
triangulate surfaces and by taking then the continuum
limit \cite{t.b.p}.

In \cite{moore1} it was argued that by construction, solutions
of ~(\ref{wdw}) obtained by conformal field theory
KPZ techniques, are wavefunctions obtained
by the insertion of local states operators.
Solutions were sought therefore, by looking for wavefunctions
of insertions of conformal operators on a disk.
In pure gravity, the insertion of the
cosmological constant is equivalent to marking a point
on  the disk.
The explicit  formula for the partition function
of the marked (punctured) disk is:
\hsm $A Z(A,l)$ \hsm where $Z(A,l)$ is given by ~(\ref{part})).
After applying ~(\ref{int}) this gives for the
marked disk wave function
\bena
\Psi_{punct. \ disk}=(\sqrt{\mu})^{\frac{1}{2}}K_{\frac{1}{2}}
(2 \sqrt{\mu}l)
\eena
As observed in \cite{moore1} this is
according to ~(\ref{solution})
a solution of the 2D WDW equation ~(\ref{wdw}).
{}From the solutions achieved by inserting local
conformal operators,
this is in fact the only one which admits a simple
geometrical picture.

By inserting the boundary operator,
in the partition function, one obtains,as in \cite{moore1}
\hsm $l Z(A,l)$.
By applying the transform in ~(\ref{int})
this leads to the wave function :
\bena\label{boundary}
\Psi_{boundary} =(\sqrt{\mu})^{\frac{3}{2}}K_{\frac{3}{2}}
(2 \sqrt{\mu}l)
\eena
which according to ~(\ref{solution}) is also a solution of
of the 2D WDW equation ~(\ref{wdw}) with $\nu = 3/2$.

The solution we present in the next section
has the same functional form
like ~(\ref{boundary})  but it is obtained
by considering the amplitude for surfaces whose
boundary consists
of two loops that coincide in one point.

This geometrical interpretation of the universe
wave function as the amplitude for a universe
originating "out of nothing" to split in tangent "circles"
is a germane to the "foam" idea of Hawking and Hartle
for the texture of the universe.
%-------------------------------------------------------%
\section{Figure-eight}
%3) The solution is 8
%The 8 diagram amplitude with fixed l1, l2 depends only on total
% boundary length l1+l2 and is equivalent to  to the
%disk with 1 marked point.
%When summed over l1,l2 with l fixed, fulfills
%WDW.
%Combinatorics results (MSc, TBP).

Our technique to calculate the "figure-eight" and other
singular geometry amplitudes  is based upon exact counting of
random triangulate surfaces \cite{tutte}. We use a recursive
equation which is obtained by expressing the change in the
partition function of the discrete surface caused by removing
one boundary edge.

While the use of the recursion equation is quite complicated in
the non-singular diagram case, in the singular diagram case,
it consists of only one term which is usually related to
an  already known regular partition function.

It is lucky that the present method is complementary
to the double limit matrix model in the sense that it
needs the results of the matrix model for certain
regular surfaces but it provides results for the singular
geometries which are unavailable in the matrix model framework.

Here we
present the continuous version of the argument chain and
refere to \cite{t.b.p,thesis} for the detailed mathematical
formulae which represent these manipulations in a rigorous
quantitative way. The additional rigor \cite{t.b.p,thesis}
is related to the fact
that while in the present exposition  we operate on
unregularized  (continuum) surfaces, in \cite{t.b.p,thesis},
the lattice provides a well defined basis for counting
possibilities.

As an example of the additional rigor
provided by the discrete treatment,
consider the following example.
In the paragraphs below we use the fact
that the number of ways in which one can split
a continuum loop of length $l$ into two segments is
proportional to $l$.
In the lattice calculation this number is
explicitly found to be $L-L_{o}+1$ where
$L$  is the loop length, in lattice spacing units $a$, $L_{o}$
is the minimal length that two loops can have in lattice
spacing units (i.e 6 in the case
of triangulate lattice, 8 for a quartic lattice etc \ldots)
and $L \geq L_{o}$.
In addition to the technical advantages, the discrete
triangulation formulation has methodological
advantages. In fact much of the mathematical
formulae expressing the amplitudes properties
reported here and in \cite{t.b.p,thesis}
were suggested to us by the direct simulation and
visualization of RTS as reported in \cite{all}.

Let us now return to the estimation of the
"figure-eight" amplitude.
This is obtained by causing the disk boundary
to assume  the singular geometry shown in  (fig.2).
What is the partition function of such a geometry ?

We first remind the reader that the loop amplitude
$w(l)$
corresponds in the matrix model to the measurable
$tr \phi^{l/a}$
which creates a hole of length $l$ in a discrete surface
with lattice spacing $a$.
Its computation is made in the matrix model with the convention
that one point of the loop is actually {\em marked}.

The unmarked amplitude
$w^{mm}(l)$ \cite{moore1}
is related to the marked loop amplitude
$w(l)$ by the relation:
\hsm $w^{mm}(l)=\frac{w(l)}{l}$  which takes symmetry
into consideration.

In order to deduce from a marked point loop amplitude the
"figure-eight" amplitude, one marks another point on the loop
and identify these two points.
However, if the length of the loops is known (i.e
$l_{1},l_{2}$) then we need to mark only one point
because the position of the second one
is determined by $l_{1}$ and $l_{2}$.
We conclude that:
\be\label{config}
Z_{fig.8}(A,l_{1},l_{2})= Z(A,l_{1}+l_{2})
\ee
and does not depend on either $l_{1}$ or $l_{2}$.

This result is more transparent in the context of triangulate surfaces
(where in fact we found it initially).
There, we have expressed by a recursive equation the partition function
of a surface with two boundaries. The recursive formula sums all
possibilities of removing one boundary edge. One of the terms
we count is this "figure-eight" which is a two-loop surface
but has the weight of the $L_{1}+L_{2}+1$ disk (fig.3).
Thus ~(\ref{config}) is written in triangulate lattice terms as:
\be
Z_{fig.8}(N,L_{1}+L_{2})=Z(N-1,L_{1}+L_{2}+1)
\ee
where $N$ is the number of triangles (fig. 3).

By taking into account all $L-5$ possibilities of dividing the $L$
($L \geq 6$) loop into two (as explained above)
we find that
\be\label{fig8}
Z_{fig.8}(N,L_{1}+L_{2}=L)=(L-5)* Z(N-1,L_{1}+L_{2}+1)
\ee
which  becomes in the continuum after applying ~(\ref{int})
(using $a*L \rightarrow l \hsm,\hsm a^2*N \rightarrow A$ \cite{david})
\be\label{ampfig8}
w_{fig.8}(l)= (\sqrt{\mu})^{\frac{3}{2}}K_{\frac{3}{2}}
(2 \sqrt{\mu} l)
\ee
and is a solution of the WDW equation.

The equivalence between a disk with two marked
boundary points and
a "figure-eight", has a further geometrical meaning
which was clarified to us by E. Witten.
In order to discribe it, one can use
fig. 4\hsm.
Consider
a cut through the equator of a nodded torus.
What we find is a continuum version of the triangulate
diagram which appears in fig 3. without
specifying the values of the two loop lengths $l_{1}$ and $l_{2}$.
Fig. 4 is equivalent, after removing the singularity,
to a disk with two marked boundary points, just as the nodded
torus is equivalent, after removing the singularity, to a
sphere with two punctures.
%---------------------------------------------------------%
\section{Singular geometries}

We can further use our knowledge to evaluate other singular
geometries.These are obtained by passing from one topology to
another. A "figure-eight" (fig. 3) is the evaluation of a two
loop surface out of a disk.
The "one handle figure-eight" (fig. 5) represents the evaluation
of a one boundary torus out of a two loop surface. Its amplitude
is the amplitude of the $l$ possible two loop surfaces of which
it is decomposed. Using the result of \cite{moore1} we have
\be
w(l_{1},l_{2})=\sqrt{l_{1}}\sqrt{l_{2}}
\frac{e^{-2\sqrt{\mu}(l_{1}+l_{2})}}{l_{1}+l_{2}}
\ee
we thus find for the "one handle figure-eight" (fig. 5):
\be
\bar{w}(l)=l e^{-2\sqrt{\mu}l}
\ee

The importance of singular geometries, as found in \cite{t.b.p},
is expressed by the role they play in the recursive
relation description of partition functions of
surfaces with more then one handle or boundary.

Indeed, once the singular geometries are under control,
one can contionously relate surfaces with different
topologies by "smoothly" deforming them
into one onother while passing through the
appropriate singular topologies.

For instance the $k$ boundaries partition function is expressed
by $k$ disk-like terms corresponding to the $k$ loops, with  additional
"figure-eight"-like terms related to a $k-1$ loop surface.
The $h$ handle and $k$ boundaries partition function is expressed
as the $k$ boundaries partition function, with additional "one
handle figure-eight"-like terms related to a surface with
$h-1$ handles and $k+2$ loops.

This leads to a general strategy for the evaluation of
surfaces:
once a disk is formed it can dynamically change into
other topology with one boundary or more. The change
is done via singular geometries.
It was drawn to our attention by N. Seiberg that using this
picture one can expect to relate amplitudes of surfaces
with various boundary topologies to the disk amplitude.
The relation between the multi-loop and disk amplitudes
was found by different techniques in \cite{moore1}.

We thank
S. Elitzur,
A. Goldberger,
E. Rabinovici,
N. Seiberg and
E. Witten
for discussions.

This work was supported in part by GIF (Germany-Israel Foundation)
and by the Israeli Academy Foundation for Fundamental Research.
%----------------------------------------------------------%
\newpage
\bebib{10}
\protect\small
\addtolength{\itemsep}{-\itemsep}

\bibitem{DeWitt}
B,S.\ DeWitt, Phys.\ Rev.\ 160 (1967) 1113.
\bibitem{fund}
S. Elitzur, A.\ Forge and E.\ Rabinovici, Phys.\ Lett.\ B289 (1992) 45.
\bibitem{moore1}
G.\ Moore , N.\ Seiberg, and M.\ Staudacher, Nucl.\ Phys.\ B362 (1991) 665.
\bibitem{kpz}
F.\ David, Mod.\ Phys.\ Lett.A3 (1988) 1651.
J.\ Distler and H.\ Kawai, Nucl.\ Phys.\ B321 (1989) 509.
\bibitem{hh}
J,B.\ Hartle  and S,W.\ Hawking, Phys.\ RevD 28 (1983) 2960.
\bibitem{moore2}
P.\ Ginsparg and G.\ Moore, YCTP-p23-92, hep-th/9304011.
\bibitem{t.b.p}
E.\ Adi and S.\ Solomon, in preparation.
\bibitem{tutte}
W,T.\ Tutte,   Can. J. Math 14 (1962) 21.
\bibitem{thesis}
E.\ Adi,  M.Sc thesis (1993).
\bibitem{all}
E. Adi et. al., Phys. Lett. B320 (1994) 227.
\bibitem{david}
F.\ David, hep-th/9303127.
\eebib
%------------------------------------------------------------%
% Figure Captions %
\newpage
\centerline{\ FIGURE CAPTIONS}
\vskip 8pt
\begin{itemize}
\item{Fig.1 \hsm A punctured disk.
}
\item{Fig.2 \hsm "Figure-eight".
}
\item{Fig.3 \hsm A triangulate "figure-eight" as a part of
combinatorial counting.In the figure ,$A_{N L_{1} L_{2}}$
is the number of possible two boundary triangulate surfaces
with $N=10 \hsm L_{1}=6 \hsm L_{2}=4$.\ $a_{N-1 L_{1}+L_{2}+1}$
is the number of disk topology triangulate surfaces composed of
$N-1=9$ triangles and boundary length of $L_{1}+L_{2}+1=11$.
The disk is the result of the removal of one boundary edge.
(The one at the top of the diagram.)
}
\item{Fig.4 \hsm A cut through the equator of
the nodded torus and the figure-eight result.
After the singularity is removed we find a disk
with two marked boundary points.
}
\item{Fig.5 \hsm One-handle figure-eight.
}
\end{itemize}
%----------------------------------------------------------%
\end{document}